\documentclass[12pt]{article}
\usepackage{amsmath,amssymb,amsthm,amsxtra,overpic,bbm,bm,epsfig,subfigure}
\usepackage{hyperref}
\usepackage{mathrsfs}
\usepackage{enumitem}
\usepackage{graphicx}
\usepackage{color}
\usepackage{comment}
\usepackage{epstopdf}
\usepackage{float}
\usepackage{cite}
\usepackage{slashed,stmaryrd}

\def\thefootnote{\fnsymbol{footnote}}

\begin{document}

\vspace{0.2cm}

\begin{center}
{\Large\bf Cosmic Flavor Hexagon for Ultrahigh-energy Neutrinos and Antineutrinos at Neutrino Telescopes}
\end{center}

\vspace{0.2cm}

\begin{center}
{\bf Shun Zhou}~\footnote{E-mail: zhoush@ihep.ac.cn}
\\
\vspace{0.2cm}
{\small
Institute of High Energy Physics, Chinese Academy of Sciences, Beijing 100049, China\\
School of Physical Sciences, University of Chinese Academy of Sciences, Beijing 100049, China\\}
\end{center}

\vspace{1.5cm}

\begin{abstract}
In this paper, we propose a hexagonal description for the flavor composition of ultrahigh-energy (UHE) neutrinos and antineutrinos, which will hopefully be determined at the future large neutrino telescopes. With such a geometrical description, we are able to clearly separate the individual flavor composition of neutrinos from that of antineutrinos in one single regular hexagon, which can be regarded as a natural generalization of the widely-used ternary plot. For illustration, we consider the $pp$ or $p\gamma$ collisions as the dominant production mechanism for UHE neutrinos and antineutrinos in the cosmic accelerator, and investigate how neutrino oscillations in the standard picture and in the presence of Lindblad decoherence could change the flavor composition of neutrinos and antineutrinos at neutrino telescopes.
\end{abstract}


\def\thefootnote{\arabic{footnote}}
\setcounter{footnote}{0}

\newpage

\section{Introduction}\label{sec:intro}

The origin of high-energy cosmic rays has been a long-standing puzzle in particle astrophysics and astronomy~\cite{Bhattacharjee:1998qc}. If ultrahigh-energy (UHE) neutrinos are produced as well in the cosmic accelerators, where the cosmic-ray protons are accelerated to extremely-high energies, the detection of such UHE neutrinos will provide an important clue to the cosmic-ray puzzle~\cite{Gaisser:1994yf, Xing:2011zza}. Neutrinos as a cosmic messenger have several advantages. First, neutrinos are electrically neutral, so their direction of motion will not be affected by the intergalactic magnetic fields and they point back directly to the location of the source. Second, neutrinos come with three flavors, bearing extra useful information about the production and acceleration mechanisms. Third, neutrinos are rarely interacting with matter, so they are not significantly absorbed during their propagation to the Earth and can be utilized to probe the source at a rather far distance.

The accelerated cosmic-ray protons are very likely to interact with ambient protons or photons in the cosmic accelerator, producing a large amount of pions in the energetic proton-proton ($pp$) or proton-photon ($p\gamma$) collisions. The subsequent decays of pions (i.e., $\pi^+ \to \mu^+ + \nu^{}_\mu$ and $\pi^- \to \mu^- + \overline{\nu}^{}_\mu$) and those of the secondary muons (i.e., $\mu^+ \to e^+ + \overline{\nu}^{}_\mu + \nu^{}_e$ and $\mu^- \to e^- + \nu^{}_\mu + \overline{\nu}^{}_e$) lead to the production of UHE cosmic neutrinos. If the $pp$ collisions are the dominant mechanism for neutrino production, the flavor composition of UHE neutrinos and antineutrinos is given by
\begin{eqnarray}\label{eq:ppratio}
f^{\rm S}_{\nu^{}_e} : f^{\rm S}_{\overline{\nu}^{}_e} : f^{\rm S}_{\nu^{}_\mu} : f^{\rm S}_{\overline{\nu}^{}_\mu} : f^{\rm S}_{\nu^{}_\tau} : f^{\rm S}_{\overline{\nu}^{}_\tau} = \frac{1}{6} : \frac{1}{6} : \frac{1}{3} : \frac{1}{3} : 0 : 0 \; ,
\label{eq:flavorSpp}
\end{eqnarray}
where the superscript ``S" denotes the flavor composition at the source, $\pi^+$'s and $\pi^-$'s are assumed to be equally generated because of the isospin conservation in the strong interaction. In contrast, if the $p\gamma$ collisions dominate over other interactions, then we have
\begin{eqnarray}\label{eq:pgratio}
f^{\rm S}_{\nu^{}_e} : f^{\rm S}_{\overline{\nu}^{}_e} : f^{\rm S}_{\nu^{}_\mu} : f^{\rm S}_{\overline{\nu}^{}_\mu} : f^{\rm S}_{\nu^{}_\tau} : f^{\rm S}_{\overline{\nu}^{}_\tau} =\frac{1}{3} : 0 : \frac{1}{3} : \frac{1}{3} : 0 : 0 \; ,
\label{eq:flavorSpg}
\end{eqnarray}
where only $\pi^+$'s are produced due to the conservation of electric charges. Though the neutrino flavor ratios at the source in Eqs.~(\ref{eq:ppratio}) and (\ref{eq:pgratio}) are by no means exact, where the multiple-pion production channels and the energy dependence should also be taken into account, it is evidently important to discriminate between neutrinos and antineutrinos in the ongoing and forthcoming neutrino telescopes in order to pin down the true production mechanism for UHE neutrinos.

It is Glashow who first suggested observing the charged weak gauge boson $W^-$ by using UHE cosmic-ray antineutrinos via the resonant production $\overline{\nu}^{}_e + e^- \to W^- \to$ anything, which is now known as the Glashow resonance (GR)~\cite{Glashow:1960zz, Berezinsky:1977sf}. For the $W^-$-boson mass $M^{}_{W} = 80.4~{\rm GeV}$ and the electron mass $m^{}_e = 0.511~{\rm MeV}$, the energy threshold for the GR to take place can be estimated as $E^{\rm th}_{\overline{\nu}^{}_e} \approx M^2_W/(2m^{}_e) \approx 6.3~{\rm PeV}$. Since the GR is exclusively sensitive to $\overline{\nu}^{}_e$, it can be used to determine whether $pp$ or $p\gamma$ collisions dominate the production of UHE neutrinos~\cite{Anchordoqui:2004eb, Bhattacharjee:2005nh, Pakvasa:2007dc, Maltoni:2008jr, Xing:2011zm, Bhattacharya:2011qu, Barger:2014iua, Nunokawa:2016pop, Biehl:2016psj, Huang:2019hgs}. Recently, there appears a good candidate for the GR event at the IceCube detector~\cite{Lu2018}. Motivated by this exciting progress, we further assume that the flavor composition of both neutrinos and antineutrinos can be ultimately determined in the long-term operation of current and future neutrino telescopes~\cite{Aartsen:2013jdh, Aartsen:2014njl, Adrian-Martinez:2016fdl}. Then an immediate question is whether there is a suitable parametrization of the flavor composition of UHE neutrinos and antineutrinos such that they can be presented clearly in a single plot. Similar to the widely-adopted ternary plot for the flavor composition of the total fluxes of neutrinos and antineutrinos, such a pictorial presentation will be useful in reporting experimental results and also in the phenomenological studies of UHE neutrinos, in which the flavor composition of neutrinos differs from that of antineutrinos~\cite{Beacom:2002vi, Beacom:2003zg, Pagliaroli:2015rca, Bustamante:2015waa, Bustamante:2016ciw, Denton:2018aml, Bustamante:2020niz, Abdullahi:2020rge}.

In this paper, we take up the task to answer this immediate question and put forward a novel geometrical presentation of the flavor composition of UHE neutrinos and antineutrinos in Sec.~\ref{sec:hexagon}. Our presentation via a regular hexagon serves as a natural generalization of the ordinary ternary plot. Moreover, two different scenarios for flavor conversions of UHE neutrinos and antineutrinos are considered in Sec.~\ref{sec:scenarios} to illustrate possible applications of the flavor hexagon. First, we reexamine the flavor composition of UHE neutrinos at the detector in the standard picture of three-flavor neutrino oscillations, in which the latest results of neutrino oscillation parameters from the global-fit analysis of neutrino oscillation data are input. Second, neutrino oscillations in the presence of Lindblad decoherence are investigated, where possible deviations of the flavor composition from the standard prediction can be observed. Finally, we summarize our main results in Sec.~\ref{sec:summary}.

\section{Cosmic Flavor Hexagon}\label{sec:hexagon}

Before introducing the cosmic flavor hexagon, we have to first explain how to parametrize the flavor composition of UHE neutrinos. As usual, the fluxes of neutrinos $\nu^{}_\alpha$ and those of antineutrinos $\overline{\nu}^{}_\alpha$ (for $\alpha = e, \mu, \tau$) are denoted as $\phi^{}_{\nu^{}_\alpha}$ and $\phi^{}_{\overline{\nu}^{}_\alpha}$, respectively. The flavor composition is defined as $f^{}_\alpha \equiv \phi^{}_\alpha/\phi^{}_0$, where $\phi^{}_\alpha \equiv \phi^{}_{\nu^{}_\alpha} + \phi^{}_{\overline{\nu}^{}_\alpha}$ is the sum of the fluxes of neutrinos $\nu^{}_\alpha$ and antineutrinos $\overline{\nu}^{}_\alpha$ for an individual flavor and $\phi^{}_0 \equiv \phi^{}_e + \phi^{}_\mu + \phi^{}_\tau$ is the total flux of neutrinos and antineutrinos of all three flavors. A convenient parametrization of neutrino flavor composition via two angles $\{\xi, \zeta\}$ has been proposed in Ref.~\cite{Xing:2006uk}, namely,
\begin{eqnarray}\label{eq:xizeta}
f^{}_e : f^{}_\mu : f^{}_\tau = \sin^2 \xi \cos^2 \zeta : \cos^2 \xi \cos^2 \zeta : \sin^2 \zeta \; ,
\end{eqnarray}
where $f^{}_\alpha \equiv f^{}_{\nu^{}_\alpha} + f^{}_{\overline{\nu}^{}_\alpha}$ is rewritten in terms of $f^{}_{\nu^{}_\alpha} \equiv \phi^{}_{\nu^{}_\alpha}/\phi^{}_\alpha$ and $f^{}_{\overline{\nu}^{}_\alpha} \equiv \phi^{}_{\overline{\nu}^{}_\alpha}/\phi^{}_\alpha$ (for $\alpha = e, \mu, \tau$). With loss of generality, the physical values of $\xi$ and $\zeta$ can be restricted into the range $[0, 90^\circ]$. For instance, the flavor ratio is $f^{}_e : f^{}_\mu : f^{}_\tau = 1/3 : 2/3 : 0$ for cosmic neutrino production via the $\pi$-$\mu$ decay chain, corresponding to $\xi = \arcsin(1/\sqrt{3}) \approx 35.3^\circ$ and $\zeta = 0$, which is applicable to both cases of $pp$ and $p\gamma$ collisions at the source. Although we shall fix $\xi = 35.3^\circ$ and $\zeta = 0$ at the source in the present work, one can scan over all possible values of $\{\xi, \zeta\}$ for the most general flavor composition at the source and explore their allowed regions at neutrino telescopes~\cite{Serpico:2005sz, Serpico:2005bs, Winter:2006ce, Cuoco:2006qd, Meloni:2006gv, Xing:2008fg}.

If neutrinos and antineutrinos of all three flavors are completely distinguishable, then we can generalize the parametrization Eq.~(\ref{eq:xizeta}) to the following one
\begin{eqnarray}\label{eq:xizetax}
f^{}_{\nu^{}_e} : f^{}_{\overline{\nu}^{}_e} : f^{}_{\nu^{}_\mu} : f^{}_{\overline{\nu}^{}_\mu} : f^{}_{\nu^{}_\tau} : f^{}_{\overline{\nu}^{}_\tau} = x^{}_e f^{}_e : (1 - x^{}_e) f^{}_e : x^{}_\mu f^{}_\mu : (1 - x^{}_\mu) f^{}_\mu : x^{}_\tau f^{}_\tau : (1 - x^{}_\tau) f^{}_\tau \; ,
\end{eqnarray}
where $x^{}_\alpha \equiv f^{}_{\nu^{}_\alpha}/f^{}_\alpha$ denotes the fraction of neutrinos in each flavor (for $\alpha = e, \mu, \tau$) and $f^{}_\alpha$ have been given in Eq.~(\ref{eq:xizeta}). It is worth mentioning that the parametrization in Eq.~(\ref{eq:xizetax}) can be applied to the flavor composition at the source as well as that at the detector. In these cases, a superscript ``S" or ``D" will be attached to the flavor composition in order to avoid confusion.
\begin{figure}[!t]
\begin{center}	
\includegraphics[width=0.7\textwidth,angle=0]{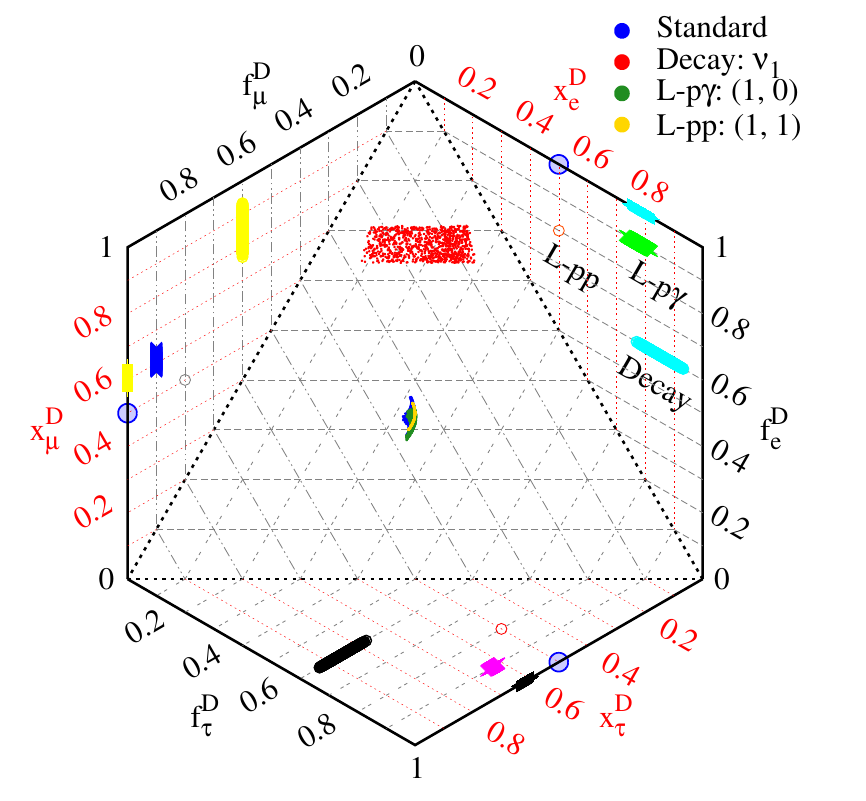}	
\caption{The cosmic flavor hexagon for UHE neutrinos and antineutrinos at the detector, where $\{f^{\rm D}_e, f^{\rm D}_\mu, f^{\rm D}_\tau\}$ denote the flavor composition of neutrinos and antineutrinos, and $\{x^{\rm D}_e, x^{\rm D}_\mu, x^{\rm D}_\tau\}$ stand for the neutrino fraction in each flavor $x^{\rm D}_\alpha \equiv f^{\rm D}_{\nu^{}_\alpha}/f^{\rm D}_\alpha$ for $\alpha = e, \mu, \tau$. Within the equilateral triangle, we show the allowed regions of the flavor composition $\{f^{\rm D}_e, f^{\rm D}_\mu, f^{\rm D}_\tau\}$ in the standard picture of neutrino oscillations (blue dots), in the case of neutrino decays with a stable $\nu^{}_1$ (red dots), with the Lindblad decoherence for $p\gamma$ collisions (labelled by ``L-p$\gamma$") at the source (green dots), and that the Lindblad decoherence for $pp$ collisions (labelled by ``L-pp") at the source (golden dots). The projection into each neutrino flavor $\{x^{\rm D}_e, x^{\rm D}_\mu, x^{\rm D}_\tau\}$ can be read off from one of two equal sides of the surrounding isosceles triangle. See the main text for the discussions on different scenarios.}\label{fig:hexagon}
\end{center}
\end{figure}

Then our primary goal in this section is to visualize the parametrization of the flavor composition in Eq.~(\ref{eq:xizetax}). At the source, we consider either the $pp$ or $p\gamma$ collisions as the dominant production mechanism for UHE neutrinos and antineutrinos. Hence the initial flavor composition in either $pp$ or $p\gamma$ case is accordingly given in Eq.~(\ref{eq:ppratio}) or Eq.~(\ref{eq:pgratio}). After their production, UHE neutrinos and antineutrinos will travel a long distance to neutrino telescopes at the Earth, during which they may change flavors. However, we postpone the discussions about neutrino flavor conversions to the next section, and for the moment explain the geometrical presentation of neutrino flavor composition at neutrino telescopes.

Without the ability to discriminate between neutrinos and antineutrinos at the detector, one usually presents the flavor composition in Eq.~(\ref{eq:xizeta}) in the ternary plot, where $f^{\rm D}_e + f^{\rm D}_\mu + f^{\rm D}_\tau = 1$ automatically holds. As shown in Fig.~\ref{fig:hexagon}, the ordinary ternary plot is localized in the center as the equilateral triangle, whose three sides are plotted as thick dotted lines. To read off the coordinate values of any point within the equilateral triangle, one should follow the grid lines of the same style to the coordinate axis, namely, the gray dashed lines for $f^{\rm D}_e$, the gray double-dotted-dashed lines for $f^{\rm D}_\mu$ and the gray double-dashed lines for $f^{\rm D}_\tau$. The basic idea here is exactly the same as for the ordinary ternary plot, except for the projections from the sides of the equilateral triangle to those of the regular hexagon.

As three sides of the regular hexagon have been occupied by $\{f^{\rm D}_e, f^{\rm D}_\mu, f^{\rm D}_\tau\}$, we can implement the remaining ones to represent $\{x^{\rm D}_e, x^{\rm D}_\mu, x^{\rm D}_\tau\}$. First, since both $x^{\rm D}_\alpha$ and $f^{\rm D}_\alpha$ refer to the same flavor, it is natural to associate them with a single isosceles triangle. In the counterclockwise order, one can find the isosceles triangles of $\{f^{\rm D}_e, x^{\rm D}_e\}$, $\{f^{\rm D}_\mu, x^{\rm D}_\mu\}$ and $\{f^{\rm D}_\tau, x^{\rm D}_\tau\}$ in Fig.~\ref{fig:hexagon}. Second, the coordinate axes of $\{x^{\rm D}_e, x^{\rm D}_\mu, x^{\rm D}_\tau\}$ have been highlighted in red. The value of $x^{\rm D}_\alpha$, i.e., the fraction of neutrinos, can be read off from the coordinate axis, while the fraction of antineutrinos is given by $1 - x^{\rm D}_\alpha$. All the lines parallel to the coordinate axis of $x^{\rm D}_\alpha$ in a single isosceles triangle can be used equally to denote the neutrino fraction, whose value is determined according to the coordinate axis. The thin dotted grid lines in red have been added in the corresponding isosceles triangle for this purpose.

To be explicit, we shall give two simple examples of the flavor conversions of UHE neutrinos and antineutrinos in the next section and apply the hexagonal description in Fig.~\ref{fig:hexagon} to show the final flavor composition at the detector. The allowed region of the flavor composition in each scenario, as shown in Fig.~\ref{fig:hexagon}, will be explained in detail.

\section{Flavor Composition at Neutrino Telescopes}\label{sec:scenarios}

\subsection{Standard Oscillations}

The first example is the standard picture of three-flavor neutrino oscillations. Since the cosmic accelerators are likely to be extragalactic, the distance between the sources and the detectors turns out to be much longer than neutrino oscillation lengths even for UHE neutrinos and antineutrinos. As a result, the oscillation terms will be averaged out and the flavor composition of UHE neutrinos and antineutrinos at the detector can be calculated immediately as
\begin{eqnarray}
f^{\rm D}_{\nu^{}_\alpha} &=& \sum_{\beta = e, \mu, \tau} P^{+}_{\alpha \beta} f^{\rm S}_{\nu^{}_\beta} = \sum_{\beta = e, \mu, \tau} \sum^3_{i = 1} |U^{}_{\alpha i}|^2 |U^{}_{\beta i}|^2 f^{\rm S}_{\nu^{}_\beta} \; , \label{eq:Ppstd}\\
f^{\rm D}_{\overline{\nu}^{}_\alpha} &=& \sum_{\beta = e, \mu, \tau} P^{-}_{\alpha \beta} f^{\rm S}_{\overline{\nu}^{}_\beta} = \sum_{\beta = e, \mu, \tau} \sum^3_{i = 1} |U^{}_{\alpha i}|^2 |U^{}_{\beta i}|^2 f^{\rm S}_{\overline{\nu}^{}_\beta} \; ,
\label{eq:Pmstd}
\end{eqnarray}
where $P^{+}_{\alpha \beta} \equiv P(\nu^{}_\beta \to \nu^{}_\alpha)$ and $P^{-}_{\alpha \beta} \equiv P(\overline{\nu}^{}_\beta \to \overline{\nu}^{}_\alpha)$ denote the oscillation probabilities  neutrinos $\nu^{}_\beta \to \nu^{}_\alpha$ for neutrinos and $\overline{\nu}^{}_\beta \to \overline{\nu}^{}_\alpha$ for antineutrinos (for $\alpha, \beta = e, \mu, \tau$), respectively. Since the oscillation terms have been averaged out, the CP-violating term is absent and the probabilities of neutrino and antineutrino oscillations are identical, i.e., $P^+_{\alpha \beta} = P^-_{\alpha \beta}$, and $P^\pm_{\alpha \beta} = P^\pm_{\beta \alpha}$ holds as well.

In the standard parametrization, the neutrino flavor mixing matrix $U$ can be explicitly written in terms of three mixing angles $\{\theta^{}_{12}, \theta^{}_{13}, \theta^{}_{23}\}$ and one Dirac-type CP-violating phase $\delta$, namely,
\begin{eqnarray}
U = \left( \begin{matrix} c^{}_{13} c^{}_{12} & c^{}_{13} s^{}_{12} & s^{}_{13} e^{-{\rm i}\delta} \cr -s^{}_{12} c^{}_{23} - c^{}_{12} s^{}_{23} s^{}_{13} e^{{\rm i}\delta} & +c^{}_{12} c^{}_{23} - s^{}_{12} s^{}_{23} s^{}_{13} e^{{\rm i}\delta} & c^{}_{13} s^{}_{23} \cr +s^{}_{12} s^{}_{23} - c^{}_{12} c^{}_{23} s^{}_{13} e^{{\rm i}\delta} & -c^{}_{12} s^{}_{23} - s^{}_{12} c^{}_{23} s^{}_{13} e^{{\rm i}\delta} & c^{}_{13} c^{}_{23} \end{matrix}\right) \; ,
\label{eq:parastd}
\end{eqnarray}
with $s^{}_{ij} \equiv \sin \theta^{}_{ij}$ and $c^{}_{ij} \equiv \cos \theta^{}_{ij}$ for $ij = 12, 13, 23$. The latest global-fit analysis of neutrino oscillation data~\cite{Capozzi:2020qhw} indicates that $\theta^{}_{23} = 45^\circ$ and $\delta = 270^\circ$ are allowed at the $3\sigma$ level, which are consistent with the predictions from a $\mu$-$\tau$ reflection symmetry of the lepton flavor structure~\cite{Xing:2015fdg, Xing:2019vks}. Following the approach in Refs.~\cite{Xing:2006xd, Xing:2012sj}, we introduce the $\mu$-$\tau$-symmetry breaking parameters
\begin{eqnarray}
\Delta^{}_1 &\equiv& |U^{}_{\mu 1}|^2 - |U^{}_{\tau 1}|^2 = (\sin^2 \theta^{}_{12} - \cos^2 \theta^{}_{12} \sin^2 \theta^{}_{13})\cos 2\theta^{}_{23} + \sin 2\theta^{}_{12} \sin 2\theta^{}_{23} \sin\theta^{}_{13} \cos \delta \; , \quad \label{eq:Delta1}\\
\Delta^{}_2 &\equiv& |U^{}_{\mu 2}|^2 - |U^{}_{\tau 2}|^2 = (\cos^2 \theta^{}_{12} - \sin^2 \theta^{}_{12} \sin^2 \theta^{}_{13})\cos 2\theta^{}_{23} - \sin 2\theta^{}_{12} \sin 2\theta^{}_{23} \sin\theta^{}_{13} \cos \delta \; , \quad \label{eq:Delta2}\\
\Delta^{}_3 &\equiv& |U^{}_{\mu 3}|^2 - |U^{}_{\tau 3}|^2 = - \cos^2\theta^{}_{13} \cos 2\theta^{}_{23} \; , \label{eq:Delta3}
\end{eqnarray}
with which one can verify that $\Delta^{}_1 + \Delta^{}_2 + \Delta^{}_3 = 0$ is valid due to the unitarity of $U$. Furthermore, we observe that $\Delta^{}_1 \leftrightarrow \Delta^{}_2$ under the transformations $\cos^2 \theta^{}_{12} \leftrightarrow \sin^2 \theta^{}_{12}$ and $\sin 2\theta^{}_{12} \to - \sin 2\theta^{}_{12}$~\cite{Zhou:2016luk}. With those $\mu$-$\tau$-symmetry breaking parameters, one can obtain
\begin{eqnarray}
\left( \begin{matrix} |U^{}_{e 1}|^2 & |U^{}_{e 2}|^2 & |U^{}_{e 3}|^2 \cr |U^{}_{\mu 1}|^2 & |U^{}_{\mu 2}|^2 & |U^{}_{\mu 3}|^2 \cr |U^{}_{\tau 1}|^2 & |U^{}_{\tau 2}|^2 & |U^{}_{\tau 3}|^2 \end{matrix} \right) &=& \frac{1}{2} \left( \begin{matrix} 2\cos^2 \theta^{}_{12} & 2\sin^2 \theta^{}_{12} & 0 \cr \sin^2 \theta^{}_{12} & \cos^2 \theta^{}_{12}  & 1 \cr \sin^2 \theta^{}_{12} & \cos^2 \theta^{}_{12} & 1 \end{matrix}\right) + \frac{\sin^2\theta^{}_{13}}{2} \left( \begin{matrix} -2\cos^2 \theta^{}_{12} & -2\sin^2 \theta^{}_{12} & 2 \cr \cos^2 \theta^{}_{12} & \sin^2 \theta^{}_{12}  & -1 \cr \cos^2 \theta^{}_{12} & \sin^2 \theta^{}_{12} & -1 \end{matrix}\right) \nonumber \\
&~& + \frac{1}{2} \left( \begin{matrix} 0 & 0 & 0 \cr +\Delta^{}_1 & +\Delta^{}_2  & +\Delta^{}_3 \cr -\Delta^{}_1 & -\Delta^{}_2 & -\Delta^{}_3 \end{matrix}\right) \; ,
\label{eq:Ualfai}
\end{eqnarray}
where both two matrices in the first line of the right-hand side respect the $\mu$-$\tau$ symmetry while the one in the second line in general does not.

Taking the best-fit values of three neutrino mixing angles and the CP-violating phase in the case of normal neutrino mass ordering, together with their $3\sigma$ uncertainties, from Ref.~\cite{Esteban:2018azc},
\begin{eqnarray}
\sin^2 \theta^{}_{12} = 0.310^{+0.040}_{-0.035} \; , ~~~ \sin^2 \theta^{}_{23} = 0.563^{+0.046}_{-0.130} \; , ~~~ \sin^2 \theta^{}_{13} = 0.02237^{+0.00198}_{-0.00193} \; , ~~~ \delta = {221^\circ}^{+136^\circ}_{-77^\circ} \; ,
\label{eq:global}
\end{eqnarray}
we find $-0.208 \lesssim \Delta^{}_1 \lesssim 0.185$, $-0.280 \lesssim \Delta^{}_2 \lesssim 0.230$ and $-0.131 \lesssim \Delta^{}_3 \lesssim 0.213$ at the $3\sigma$ level. Their best-fit values are found to be $\Delta^{}_1 = - 0.141$, $\Delta^{}_2 = 0.018$, and $\Delta^{}_3 = 0.123$, implying that the partial $\mu$-$\tau$ symmetry~\cite{Xing:2014zka} with $\Delta^{}_2 = 0$ but $\Delta^{}_1 = - \Delta^{}_3 \neq 0$ is currently favored by neutrino oscillation data. If $\Delta^{}_2 = 0$ holds exactly, with the help of Eq.~(\ref{eq:Delta2}), then one can establish the relationship between the CP-violating phase and three neutrino mixing angles~\cite{Xing:2014zka}
\begin{eqnarray}
\cos \delta =  \frac{\cos^2 \theta^{}_{12} - \sin^2 \theta^{}_{12} \sin^2 \theta^{}_{13}}{\sin 2\theta^{}_{12} \tan 2\theta^{}_{23} \sin\theta^{}_{13}} \; .
\label{eq:cosdelta1}
\end{eqnarray}
When the best-fit values of three mixing angles in Eq.~(\ref{eq:global}) are input, the relation in Eq.~(\ref{eq:cosdelta1}) gives rise to $\delta \approx 129^\circ$. That both $\cos 2\theta^{}_{23}$ and $\cos \delta$ are negative leads to a remarkable cancellation between those two terms on the right-hand side of Eq.~(\ref{eq:Delta2}).
Notice that the absolute values of  $\Delta^{}_i$ (for $i = 1, 2, 3$) may not be very small (e.g., they can actually be larger than $0.2$), so they will be used only as a convenient parametrization of $|U^{}_{\alpha i}|^2$ instead of perturbation parameters. However, they are indeed excellent perturbation parameters in the $\mu$-$\tau$-symmetric limit of $\cos 2\theta^{}_{23} \to 0$ and $\cos \delta \to 0$. If the $\mu$-$\tau$ symmetry is further confirmed by more precise measurements in future neutrino oscillation experiments, one can safely expand the relevant formulas in terms of $\Delta^{}_i$ (for $i = 1, 2, 3$). In the present work, we shall stick to the exact formulas without any approximations in the following discussions.

We proceed to compute explicitly the flavor composition of neutrinos and antineutrinos at the detector. Given the initial flavor composition in Eq.~(\ref{eq:flavorSpp}) at the source in the case of $pp$ collisions, one can make use of Eqs.~(\ref{eq:Ppstd}) and (\ref{eq:Pmstd}) to obtain~\cite{Xing:2012sj}
\begin{eqnarray}
f^{\rm D}_{\nu^{}_e} = f^{\rm D}_{\overline{\nu}^{}_e} &=& \frac{1}{6} + \frac{1}{6} \sum^{}_i |U^{}_{ei}|^2 \Delta^{}_i = \frac{1}{6} (1 - 2\Delta) \; , \label{eq:estdpp}\\
f^{\rm D}_{\nu^{}_\mu} = f^{\rm D}_{\overline{\nu}^{}_\mu} &=& \frac{1}{6} + \frac{1}{6} \sum^{}_i |U^{}_{\mu i}|^2 \Delta^{}_i = \frac{1}{6} \left(1 + \Delta + \overline{\Delta} \right) \; , \label{eq:mstdpp}\\
f^{\rm D}_{\nu^{}_\tau} = f^{\rm D}_{\overline{\nu}^{}_\tau} &=& \frac{1}{6} + \frac{1}{6} \sum^{}_i |U^{}_{\tau i}|^2 \Delta^{}_i = \frac{1}{6} \left( 1 + \Delta - \overline{\Delta} \right)\; ,
\label{eq:tstdpp}
\end{eqnarray}
where $\Delta \equiv -(\cos^2 \theta^{}_{13} \cos^2 \theta^{}_{12} \Delta^{}_1 + \cos^2 \theta^{}_{13} \sin^2 \theta^{}_{12} \Delta^{}_2 + \sin^2 \theta^{}_{13} \Delta^{}_3)/2$ and $\overline{\Delta} \equiv (\Delta^2_1 + \Delta^2_2 + \Delta^2_3)/2$ have been defined and Eq.~(\ref{eq:Ualfai}) has been used. When the global-fit results of all relevant mixing parameters in Eq.~(\ref{eq:global}) are taken into account, we have $-0.0521 \lesssim \Delta \lesssim 0.0663$ and $0 \lesssim \overline{\Delta} \lesssim 0.0626$ at the $3\sigma$ level, together with the best-fit values $\Delta = 0.0434$ and $\overline{\Delta} = 0.0176$. Notice that the results in Eqs.~(\ref{eq:estdpp})-(\ref{eq:tstdpp}) are exact without any approximations in the mixing parameters. As observed in Refs.~\cite{Xing:2006xd, Xing:2012sj}, the deviation of $f^{\rm D}_e : f^{\rm D}_\mu : f^{\rm D}_\tau = (1 - 2\Delta)/3 : (1 + \Delta + \overline{\Delta})/3 : (1 + \Delta - \overline{\Delta})/3$ from the democratic flavor ratio $f^{\rm D}_e : f^{\rm D}_\mu : f^{\rm D}_\tau = 1/3 : 1/3 : 1/3$ could be at the level of $10\%$. This is still true in consideration of $3\sigma$ uncertainties in neutrino oscillation parameters from the latest global-fit analysis of neutrino oscillation data.

In the case where $p\gamma$ collisions are dominant and the initial flavor ratio is given in Eq.~(\ref{eq:flavorSpg}), one can get the flavor composition of UHE neutrinos at neutrino telescopes
\begin{eqnarray}
f^{\rm D}_{\nu^{}_e} &=& \frac{1}{3} - \frac{1}{3} \sum^{}_i |U^{}_{ei}|^2 |U^{}_{\tau i}|^2 = \frac{1}{3} \left(1 - \Xi - \Delta \right)\; , \label{eq:enepgstd} \\
f^{\rm D}_{\nu^{}_\mu} &=& \frac{1}{3} - \frac{1}{3} \sum^{}_i |U^{}_{\mu i}|^2 |U^{}_{\tau i}|^2 =  \frac{1}{6} \left(1 + \Xi + \overline{\Delta} \right)\; , \label{eq:mnepgstd}  \\
f^{\rm D}_{\nu^{}_\tau} &=& \frac{1}{3} - \frac{1}{3} \sum^{}_i |U^{}_{\tau i}|^2 |U^{}_{\tau i}|^2  = \frac{1}{6} \left(1 + \Xi + 2\Delta - \overline{\Delta} \right)\; , \label{eq:tnepgstd}
\end{eqnarray}
with $\Xi \equiv \left(\sin^2 \theta^{}_{13} + \sin^2 \theta^{}_{12} \cos^2 \theta^{}_{12} \cos^2 \theta^{}_{13}\right) \cos^2\theta^{}_{13}$, and that of UHE antineutrinos
\begin{eqnarray}
f^{\rm D}_{\overline{\nu}^{}_e} &=& \frac{1}{3} \sum^{}_i |U^{}_{ei}|^2 |U^{}_{\mu i}|^2 = \frac{1}{3} \left(\Xi - \Delta \right)\; , \label{eq:enbpgstd} \\
f^{\rm D}_{\overline{\nu}^{}_\mu} &=& \frac{1}{3} \sum^{}_i |U^{}_{\mu i}|^2 |U^{}_{\mu i}|^2 = \frac{1}{6} \left(1 - \Xi + 2\Delta + \overline{\Delta} \right)\; , \label{eq:mnbpgstd}  \\
f^{\rm D}_{\overline{\nu}^{}_\tau} &=& \frac{1}{3} \sum^{}_i |U^{}_{\tau i}|^2 |U^{}_{\mu i}|^2  = \frac{1}{6} \left(1 - \Xi - \overline{\Delta} \right)\; . \label{eq:tnbpgstd}
\end{eqnarray}
Given the values of $\sin^2\theta^{}_{12}$ and $\sin^2\theta^{}_{13}$ in Eq.~(\ref{eq:global}), one can obtain $0.213 \lesssim \Xi \lesssim 0.239$ at the $3\sigma$ level and $\Xi = 0.226$ as the best-fit value. If neutrinos and antineutrinos are indistinguishable at neutrino telescopes, only the flavor ratio $f^{\rm D}_e : f^{\rm D}_{\mu} : f^{\rm D}_{\tau} = (1 - 2\Delta)/3 : (1 + \Delta + \overline{\Delta})/3 : (1 + \Delta - \overline{\Delta})/3$ is relevant. This flavor ratio is applicable to both cases of $pp$ and $p\gamma$ collisions.

Finally, we apply the hexagonal parametrization to the flavor composition of UHE neutrinos and antineutrinos at neutrino telescopes. For both $pp$ and $p\gamma$ collisions, we have $f^{\rm D}_e : f^{\rm D}_{\mu} : f^{\rm D}_{\tau} = (1 - 2\Delta)/3 : (1 + \Delta + \overline{\Delta})/3 : (1 + \Delta - \overline{\Delta})/3$, so we focus just on the neutrino fraction $x^{\rm D}_\alpha$ for each flavor. The result is simple in the $pp$ case, i.e., $x^{\rm D}_e(pp) = x^{\rm D}_\mu(pp) = x^{\rm D}_\tau(pp) = 1/2$, as indicated in Eqs.~(\ref{eq:estdpp})-(\ref{eq:tstdpp}). In the $p\gamma$ case, we have
\begin{eqnarray}
x^{\rm D}_e(p\gamma) &=& \frac{1 - \Xi - \Delta}{1 - 2 \Delta} \; , \label{eq:xepg}\\
x^{\rm D}_\mu(p\gamma) &=& \frac{1 + \Xi + \overline{\Delta}}{2(1 + \Delta + \overline{\Delta})} \; , \label{eq:xmpg}\\
x^{\rm D}_\tau(p\gamma) &=& \frac{1 + \Xi + 2 \Delta - \overline{\Delta}}{2(1 + \Delta - \overline{\Delta})} \; , \label{eq:xtpg}
\end{eqnarray}
where the neutrino and antineutrino flavor composition in Eqs.~(\ref{eq:enepgstd})-(\ref{eq:tnbpgstd}) have been used. In view of the smallness of the best-fit values $\Delta = 0.0434$ and $\overline{\Delta} = 0.0176$, we get $x^{\rm D}_e(p\gamma) \approx 1 - \Xi \approx 0.774$ and $x^{\rm D}_\mu(p\gamma) \approx x^{\rm D}_\tau(p\gamma) \approx (1 + \Xi)/2 \approx 0.613$ as excellent approximations.

In Fig.~\ref{fig:hexagon}, the flavor composition of neutrino and antineutrinos at the detector is shown for the standard picture of three-flavor neutrino oscillations, where the $3\sigma$ ranges of three neutrino mixing angles $\{\theta^{}_{12}, \theta^{}_{13}, \theta^{}_{23}\}$ and the CP-violating phase $\delta$ are taken as input. As we have mentioned, the flavor composition $\{f^{\rm D}_e, f^{\rm D}_\mu, f^{\rm D}_\tau\}$ remains the same for both $pp$ and $p\gamma$ collisions at the source, for which the initial flavor composition is given in Eqs.~(\ref{eq:ppratio}) and (\ref{eq:pgratio}). The allowed region of $\{f^{\rm D}_e, f^{\rm D}_\mu, f^{\rm D}_\tau\}$ at the $3\sigma$ level has been presented as blue dots within the central equilateral triangle. When neutrinos and antineutrinos are completely distinguishable in the detector, the values of $\{x^{\rm D}_e, x^{\rm D}_\mu, x^{\rm D}_\tau\}$ will be relevant. In the $pp$ case, we have $x^{\rm D}_e(pp) = x^{\rm D}_\mu(pp) = x^{\rm D}_\tau(pp) = 1/2$, which have been plotted as the filled blue circles, only the centers of which are physically meaningful and the radii are chosen to make them visible. In the $p\gamma$ case, the allowed ranges of $x^{\rm D}_e(p\gamma) \in [0.739. 0.831]$, $x^{\rm D}_\mu(p\gamma) \in [0.567, 0.646]$ and $x^{\rm D}_\tau(p\gamma) \in [0.588, 0.648]$ at the $3\sigma$ level are plotted as cyan, yellow and black ``+" points along the $\{x^{\rm D}_e, x^{\rm D}_\mu, x^{\rm D}_\tau\}$ axes. As emphasized in Ref.~\cite{Xing:2011zm}, the GR events at neutrino telescopes will be very sensitive to the fraction of $\overline{\nu}^{}_e$ characterized by $1 - x^{\rm D}_e$, making it promising to discriminate $pp$ from $p\gamma$ collisions.

\subsection{Lindblad Decoherence}

As an illustrative example for the nonstandard scenario of the flavor conversion of UHE neutrinos and antineutrinos, we implement the Lindblad equation to describe environmental impact on the time evolution of neutrino or antineutrino states. In the density matrix formulation of quantum mechanics, the Lindblad equation is given by~\cite{Lindblad:1975ef, Gorini:1975nb}
\begin{eqnarray}
\frac{{\rm d}\rho(t)}{{\rm d}t} = - {\rm i} [{\cal H}, \rho(t)] + {\cal L}[\rho(t)] \; ,
\label{eq:lindblad}
\end{eqnarray}
where $\rho(t)$ and ${\cal H}$ stand for the density matrix and the effective Hamiltonian of the neutrino system, respectively, and the Lindblad term ${\cal L}[\rho(t)]$ for the $N$-level system can be written as~\cite{Lindblad:1975ef, Gorini:1975nb}
\begin{eqnarray}
{\cal L}[\rho(t)] = - \frac{1}{2} \sum^{N^2-1}_{j = 1} \left[ A^\dagger_j A^{}_j \rho(t) + \rho(t) A^\dagger_j A^{}_j \right] + \sum^{N^2-1}_{j = 1} A^{}_j \rho(t) A^\dagger_j \; ,
\end{eqnarray}
with $A^{}_j$ (for $j = 1, 2, \cdots, N^2 - 1$) being a complete set of bounded operators. Obviously, the Lindblad operators characterize the dissipative effects on the subsystem described by the density matrix $\rho(t)$. The applications of the Lindblad equation to neutrino oscillations with a Hermitian~\cite{Lisi:2000zt, Benatti:2000ph, Gago:2000qc, Gago:2000nv, Ohlsson:2000mj, Benatti:2001fa, Gago:2002na, Oliveira:2014jsa, Coloma:2018idr, Gomes:2018inp, Gomes:2020muc} or a non-Hermitian effective Hamiltonian~\cite{Ohlsson:2020gxx} can be found in the literature.

For three-flavor neutrino oscillations, the most general form of Eq.~(\ref{eq:lindblad}) contains too many unknown parameters arising from the Lindblad operators, when the interactions of neutrinos or antineutrinos with the environment are not specified. It is convenient to expand the density matrix $\rho(t) = \rho^{}_0 \lambda^{}_0/\sqrt{6} + \rho^{}_i \lambda^{}_i/2$, the effective Hamiltonian ${\cal H} = B^{}_0 \lambda^{}_0/\sqrt{6} + B^{}_i \lambda^{}_i/2$, and the Lindblad operators $A^{}_j = A^j_{0} \lambda^{}_0/\sqrt{6} + A^{j}_{i} \lambda^{}_i/2$ in the basis of eight Gell-Mann matrices $\lambda^{}_i$ (for $i = 1, 2, \cdots, 8$) and the $3\times 3$ identity matrix $\lambda^{}_0$, where the summation over $\mu = 0, 1, \cdots, 8$ is implied. In this way, the Lindblad equation can be recast into the equation of motion for eight components $\rho^{}_i(t)$ of the density matrix $\rho(t)$, i.e.,
\begin{eqnarray}
\frac{{\rm d}\rho^{}_i}{{\rm d}t} = \sum^8_{j,k = 1} f^{}_{ijk} B^{}_j \rho^{}_k + \sum^8_{j = 1} \gamma^{}_{ij} \rho^{}_j \; ,
\label{eq:rhoi}
\end{eqnarray}
where $f^{}_{ijk}$ denotes the ${\rm SU}(3)$ structure constants and $\gamma^{}_{ij} = - \delta^{}_{ij} \sum_k {\bf a}^{}_k \cdot {\bf a}^{}_k/4 + {\bf a}^{}_i \cdot {\bf a}^{}_j/12$ with the eight-dimensional vector ${\bf a}^{}_k \equiv (A^{1}_{k}, A^{2}_{k}, \cdots, A^{8}_{k})$~\cite{Gomes:2020muc}. Under the requirement for an increasing von Neumann entropy and the probability conservation in the subsystem, we have $A^\dagger_j = A^{}_j$ and $\rho^{}_0(t) = \sqrt{2/3}$ that is time independent~\cite{Benatti:2000ph}. If the simple choice of a diagonal form $\gamma^{}_{ij} = -\delta^{}_{ij} \gamma^{}_j$ is adopted, where the complete positivity requires $\gamma^{}_i > 0$ (for $i = 1, 2, \cdots, 8$), then the oscillation probabilities of UHE neutrinos are~\cite{Gago:2002na, Gomes:2020muc}
\begin{eqnarray}
P^{}_{\alpha \beta} = \sum^3_{i = 1} |U^{}_{\alpha i}|^2 |U^{}_{\beta i}|^2 &+& 2 \sum_{i < j} {\rm Re} \left(U^{}_{\alpha j} U^*_{\alpha i} U^*_{\beta j} U^{}_{\beta i}\right) \left(1 - e^{- \Gamma^{}_{ji} L}\right)  \nonumber \\
&-& \frac{1}{6} \left(1 - 3|U^{}_{\alpha 3}|^2\right) \left( 1 - 3 |U^{}_{\beta 3}|^2\right)  \left(1 - e^{-\gamma^{}_8 L}\right) \nonumber \\
&-& \frac{1}{2} \left(|U^{}_{\alpha 1}|^2 - |U^{}_{\alpha 2}|^2\right) \left( |U^{}_{\beta 1}|^2 - |U^{}_{\beta 2}|^2\right)  \left(1 - e^{-\gamma^{}_3 L}\right) \; ,
\label{eq:probnew}
\end{eqnarray}
where $\Gamma^{}_{21} \equiv (\gamma^{}_1 + \gamma^{}_2)/2$, $\Gamma^{}_{31} \equiv (\gamma^{}_4 + \gamma^{}_5)/2$ and $\Gamma^{}_{32} \equiv (\gamma^{}_6 + \gamma^{}_7)/2$ are all positive parameters, and $L$ is the distance between the source and the detector. When the Lindblad term is absent, namely, $\Gamma^{}_{ji} = 0$ and $\gamma^{}_3 = \gamma^{}_8 = 0$, the standard oscillation probabilities are recovered. Though the Lindblad parameters $\gamma^{}_i$'s lead to damping terms, one can easily verify that $P^{}_{\alpha e} + P^{}_{\alpha \mu} + P^{}_{\alpha \tau} = 1$ holds for each neutrino flavor $\nu^{}_\alpha$ (for $\alpha = e, \mu, \tau$).

To further simplify our discussions, we assume that only $\gamma^{}_3$ for neutrinos is nonzero. For the oscillations of UHE antineutrinos, it is reasonable to expect that the environmental effects are different from those for neutrinos, which could be a consequence of either distinguishable neutrino and antineutrino interactions or a CPT-asymmetric background. More explicitly, we have
\begin{eqnarray}
P^{\pm}_{\alpha \beta} = \sum^3_{i = 1} |U^{}_{\alpha i}|^2 |U^{}_{\beta i}|^2 - \frac{\epsilon^{}_\pm}{2} \left(|U^{}_{\alpha 1}|^2 - |U^{}_{\alpha 2}|^2\right) \left( |U^{}_{\beta 1}|^2 - |U^{}_{\beta 2}|^2\right) \; ,
\label{eq:probpm}
\end{eqnarray}
where ``$\pm$" refer to neutrinos and antineutrinos with $\epsilon^{}_+ = 1 - e^{-\gamma^{}_3 L}$ and $\epsilon^{}_- = 1 - e^{-\overline{\gamma}^{}_3 L}$, respectively. Note that the new parameters $\epsilon^{}_\pm \in [0, 1]$ depend on the magnitude of original Lindblad parameters $\gamma^{}_3$ and $\overline{\gamma}^{}_3$ and the distance $L$. The present bounds on the scale of $\gamma^{}_{ij}$ have been derived from various neutrino oscillation experiments, and the most stringent one is ${\cal O}(\gamma^{}_{ij}) < 10^{-25}~{\rm GeV}$ if no dependence of $\gamma^{}_{ij}$ on the neutrino energy is assumed~\cite{Gomes:2018inp, Gomes:2020muc}. However, for a typical distance for extragalactic sources of UHE neutrinos, we have $L = 1~{\rm Mpc} = 1.6\times 10^{38}~{\rm GeV}^{-1}$ such that $\gamma^{}_3$ or $\overline{\gamma}^{}_3$ on the order of $10^{-25}~{\rm GeV}$ results in sizable values $\epsilon^{}_\pm \approx 1$. In other words, the precision measurements of oscillation probabilities for UHE neutrinos and antineutrinos at neutrino telescopes will be extremely sensitive to the Lindblad parameters, e.g., $\gamma^{}_3$ and $\overline{\gamma}^{}_3$ in our case.

Now it is straightforward to calculate the flavor composition of neutrinos and antineutrinos at neutrino telescopes, given the initial value at the source with $pp$ or $p\gamma$ collisions. If the $pp$ collisions are the dominant mechanism for the generation of UHE neutrinos and antineutrinos, then the flavor composition at the detector is given by
\begin{eqnarray}
f^{\rm D}_{\nu^{}_\alpha} &=& \frac{1}{6} \sum_i |U^{}_{\alpha i}|^2 (1 + \Delta^{}_i) - \frac{\epsilon^{}_+}{12} \left(|U^{}_{\alpha 1}|^2 - |U^{}_{\alpha 2}|^2\right) \left(\Delta^{}_1 - \Delta^{}_2\right) \; , \label{eq:fnuLindpp} \\
f^{\rm D}_{\overline{\nu}^{}_\alpha} &=& \frac{1}{6} \sum_i |U^{}_{\alpha i}|^2 (1 + \Delta^{}_i) - \frac{\epsilon^{}_-}{12} \left(|U^{}_{\alpha 1}|^2 - |U^{}_{\alpha 2}|^2\right) \left(\Delta^{}_1 - \Delta^{}_2\right)  \label{eq:fnbLindpp}\; .
\end{eqnarray}
It is worthwhile to mention that the Lindblad decoherence terms in Eqs.~(\ref{eq:fnuLindpp}) and (\ref{eq:fnbLindpp}) vanish in the limit of $\Delta^{}_1 = \Delta^{}_2$, which could be reached for
\begin{eqnarray}
\cos\delta = \frac{1 + \sin^2 \theta^{}_{13}}{2 \tan 2\theta^{}_{23} \tan 2\theta^{}_{12} \sin \theta^{}_{13}} \; ,
\label{eq:cosdelta}
\end{eqnarray}
if $\theta^{}_{23} \neq 45^\circ$ holds. For $\theta^{}_{23} = 45^\circ$, we get $\delta = 90^\circ$ or $270^\circ$, namely, the $\mu$-$\tau$-symmetric limit where $\Delta^{}_1 = \Delta^{}_2 = 0$. Taking the best-fit values $\sin^2 \theta^{}_{12} = 0.310$, $\sin^2 \theta^{}_{23} = 0.563$ and $\sin^2 \theta^{}_{13} = 0.02237$ from Eq.~(\ref{eq:global}), one can figure out $\delta \approx 100^\circ$, which is lying outside the $3\sigma$ range of $\delta$ from the global-fit analysis~\cite{Esteban:2018azc}. As we have seen in the previous subsection, the best-fit values of all neutrino mixing parameters imply $\Delta^{}_1 \approx -0.141$ and $\Delta^{}_2 \approx 0.018$ and thus $\Delta^{}_1 = \Delta^{}_2$ seems not to be the case in nature. From Eq.~(\ref{eq:fnuLindpp}), one obtains
\begin{eqnarray}
f^{\rm D}_{\nu^{}_e} &=& \frac{1}{6} \left(1 - 2\Delta\right) - \frac{\epsilon^{}_+}{12}\cos^2 \theta^{}_{13} \cos 2\theta^{}_{12} \left(\Delta^{}_1 - \Delta^{}_2\right) \; , \\
f^{\rm D}_{\nu^{}_\mu} &=& \frac{1}{6} \left(1 + \Delta + \overline{\Delta}\right) + \frac{\epsilon^{}_+}{24} \left[\cos^2 \theta^{}_{13} \cos 2\theta^{}_{12} \left(\Delta^{}_1 - \Delta^{}_2\right) - \left(\Delta^{}_1 - \Delta^{}_2\right)^2\right] \; , \\
f^{\rm D}_{\nu^{}_\tau} &=& \frac{1}{6} \left(1 + \Delta - \overline{\Delta}\right) + \frac{\epsilon^{}_+}{24} \left[\cos^2 \theta^{}_{13} \cos 2\theta^{}_{12} \left(\Delta^{}_1 - \Delta^{}_2\right) + \left(\Delta^{}_1 - \Delta^{}_2\right)^2\right] \; ,
\end{eqnarray}
for the neutrino flavor composition, while the same formulas can be carried over for antineutrinos but with $\epsilon^{}_+$ replaced by $\epsilon^{}_-$. Furthermore, the neutrino fractions are found to be
\begin{eqnarray}
x^{\rm D}_e(pp) &=&  \frac{2(1 - 2\Delta) - \epsilon^{}_+ \cos^2
\theta^{}_{13} \cos 2\theta^{}_{12} (\Delta^{}_1 - \Delta^{}_2)}{4(1 - 2\Delta) - (\epsilon^{}_+ + \epsilon^{}_-)\cos^2
\theta^{}_{13} \cos 2\theta^{}_{12} (\Delta^{}_1 - \Delta^{}_2)} \; , \label{eq:xeppLind}\\
x^{\rm D}_\mu(pp) &=&  \frac{4(1 + \Delta + \overline{\Delta}) + \epsilon^{}_+ \left[\cos^2 \theta^{}_{13} \cos 2\theta^{}_{12} (\Delta^{}_1 - \Delta^{}_2) - (\Delta^{}_1 - \Delta^{}_2)^2\right]}{8(1  + \Delta + \overline{\Delta}) + (\epsilon^{}_+ + \epsilon^{}_-) \left[\cos^2 \theta^{}_{13} \cos 2\theta^{}_{12} (\Delta^{}_1 - \Delta^{}_2) - (\Delta^{}_1 - \Delta^{}_2)^2 \right]} \; , \label{eq:xmppLind}\\
x^{\rm D}_\tau(pp) &=&  \frac{4(1 + \Delta - \overline{\Delta}) + \epsilon^{}_+ \left[\cos^2 \theta^{}_{13} \cos 2\theta^{}_{12} (\Delta^{}_1 - \Delta^{}_2) + (\Delta^{}_1 - \Delta^{}_2)^2\right]}{8(1  + \Delta - \overline{\Delta}) + (\epsilon^{}_+ + \epsilon^{}_-) \left[\cos^2 \theta^{}_{13} \cos 2\theta^{}_{12} (\Delta^{}_1 - \Delta^{}_2) + (\Delta^{}_1 - \Delta^{}_2)^2 \right]} \; . \label{eq:xtppLind}
\end{eqnarray}
By setting the Lindblad parameters $\epsilon^{}_\pm$ to zero or $\epsilon^{}_+ = \epsilon^{}_- \neq 0$, we come back to $x^{\rm D}_e(pp) = x^{\rm D}_\mu(pp) = x^{\rm D}_\tau(pp) = 1/2$ as in the scenario of standard neutrino oscillations. If the $p\gamma$ collisions are dominant at the source, the initial flavor composition is given in Eq.~(\ref{eq:pgratio}) and the deviations of the final flavor composition at the detector from those in Eqs.~(\ref{eq:enepgstd})-(\ref{eq:tnbpgstd}) in the standard scenario read
\begin{eqnarray}
\delta f^{\rm D}_{\nu^{}_\alpha} &=&  + \frac{\epsilon^{}_+}{6} \left(|U^{}_{\alpha 1}|^2 - |U^{}_{\alpha 2}|^2\right) \left(|U^{}_{\tau 1}|^2 - |U^{}_{\tau 2}|^2\right) \; , \label{eq:fnuLindpg} \\
\delta f^{\rm D}_{\overline{\nu}^{}_\alpha} &=& - \frac{\epsilon^{}_-}{6} \left(|U^{}_{\alpha 1}|^2 - |U^{}_{\alpha 2}|^2\right) \left(|U^{}_{\mu 1}|^2 - |U^{}_{\mu 2}|^2\right) \; . \label{eq:fnbLindpg}
\end{eqnarray}
More explicitly, the deviations of the neutrino flavor composition can be written as
\begin{eqnarray}
\delta f^{\rm D}_{\nu^{}_e} &=& - \frac{\epsilon^{}_+}{12} \cos^2 \theta^{}_{13} \cos 2\theta^{}_{12} \left[ \cos^2 \theta^{}_{13} \cos 2\theta^{}_{12} + (\Delta^{}_1 - \Delta^{}_2)\right] \; , \\
\delta f^{\rm D}_{\nu^{}_\mu} &=& + \frac{\epsilon^{}_+}{24} \left[ \cos^4 \theta^{}_{13} \cos^2 2\theta^{}_{12} - (\Delta^{}_1 - \Delta^{}_2)^2\right] \; , \\
\delta f^{\rm D}_{\nu^{}_\tau} &=& + \frac{\epsilon^{}_+}{24} \left[ \cos^4 \theta^{}_{13} \cos^2 2\theta^{}_{12} + 2 \cos^2 \theta^{}_{13} \cos 2\theta^{}_{12}(\Delta^{}_1 - \Delta^{}_2) + (\Delta^{}_1 - \Delta^{}_2)^2\right] \; ;
\end{eqnarray}
while those for antineutrinos are
\begin{eqnarray}
\delta f^{\rm D}_{\overline{\nu}^{}_e} &=& + \frac{\epsilon^{}_-}{12} \cos^2 \theta^{}_{13} \cos 2\theta^{}_{12} \left[ \cos^2 \theta^{}_{13} \cos 2\theta^{}_{12} - (\Delta^{}_1 - \Delta^{}_2)\right] \; , \\
\delta f^{\rm D}_{\overline{\nu}^{}_\mu} &=& - \frac{\epsilon^{}_-}{24} \left[ \cos^4 \theta^{}_{13} \cos^2 2\theta^{}_{12} - 2 \cos^2 \theta^{}_{13} \cos 2\theta^{}_{12}(\Delta^{}_1 - \Delta^{}_2) + (\Delta^{}_1 - \Delta^{}_2)^2\right] \; , \\
\delta f^{\rm D}_{\overline{\nu}^{}_\tau} &=& - \frac{\epsilon^{}_-}{24} \left[ \cos^4 \theta^{}_{13} \cos^2 2\theta^{}_{12} - (\Delta^{}_1 - \Delta^{}_2)^2\right]\; .
\end{eqnarray}
In order to see how the Lindblad decoherence modifies the flavor composition in the $p\gamma$ case, we derive the analytical formulas of $x^{\rm D}_\alpha(p\gamma)$ (for $\alpha = e, \mu, \tau$) by making some approximations. As one can observe from the expressions of $\delta f^{\rm D}_{\nu^{}_\alpha}$ and $\delta f^{\rm D}_{\overline{\nu}^{}_\alpha}$, the subleading terms are all proportional to $\Delta^{}_1 - \Delta^{}_2$, which is much smaller than the leading terms and can be ignored in the estimation. Consequently, the neutrino fractions in Eqs.~(\ref{eq:xepg})-(\ref{eq:xtpg}) will be modified to be
\begin{eqnarray}
x^{\rm D}_e(p\gamma) &\approx& \frac{1 - \Xi - \Delta - \epsilon^{}_+ \cos^4 \theta^{}_{13} \cos^2 2\theta^{}_{12}/4}{1 - 2 \Delta - (\epsilon^{}_+ - \epsilon^{}_-) \cos^4 \theta^{}_{13} \cos^2 2\theta^{}_{12}/4} \; , \label{eq:xepgLind}\\
x^{\rm D}_\mu(p\gamma) &\approx& \frac{1 + \Xi + \overline{\Delta} + \epsilon^{}_+ \cos^4 \theta^{}_{13} \cos^2 2\theta^{}_{12}/4}{2(1 + \Delta + \overline{\Delta}) + (\epsilon^{}_+ - \epsilon^{}_-) \cos^4 \theta^{}_{13} \cos^2 2\theta^{}_{12}/4} \; , \label{eq:xmpgLind}\\
x^{\rm D}_\tau(p\gamma) &\approx& \frac{1 + \Xi + 2 \Delta - \overline{\Delta} + \epsilon^{}_+ \cos^4 \theta^{}_{13} \cos^2 2\theta^{}_{12}/4}{2(1 + \Delta - \overline{\Delta}) + (\epsilon^{}_+ - \epsilon^{}_-) \cos^4 \theta^{}_{13} \cos^2 2\theta^{}_{12}/4} \; . \label{eq:xtpgLind}
\end{eqnarray}
Hence we obtain the neutrino fractions $\{x^{\rm D}_e, x^{\rm D}_\mu, x^{\rm D}_\tau\}$ in Eqs.~(\ref{eq:xeppLind})-(\ref{eq:xtppLind}) for the $pp$ collisions and those in Eqs.~(\ref{eq:xepgLind})-(\ref{eq:xtpgLind}) for the $p\gamma$ collisions. 

For illustration, we first fix the parameters $\epsilon^{}_\pm = 1$ in the Lindblad term for both $pp$ and $p\gamma$ cases, but one can examine the impact of different values of $\epsilon^{}_\pm$ in a similar way. It is straightforward to verify that $\{f^{\rm D}_e, f^{\rm D}_\mu, f^{\rm D}_\tau\}$ take the same values in these two cases, as $\delta f^{\rm D}_{\nu^{}_\alpha}$ in Eq.~(\ref{eq:fnuLindpg}) and $\delta f^{\rm D}_{\overline{\nu}^{}_\alpha}$ in Eq.~(\ref{eq:fnbLindpg}) will cancel with each other for $\epsilon^{}_+ = \epsilon^{}_-$, leaving only the terms proportional to $\Delta^{}_1 - \Delta^{}_2$. In Fig.~\ref{fig:hexagon}, we show the allowed region of $\{f^{\rm D}_e, f^{\rm D}_\mu, f^{\rm D}_\tau\}$ at the $3\sigma$ level as golden dots in the $pp$ case with $(\epsilon^{}_+, \epsilon^{}_-) = (1, 1)$, which coincides exactly with that in the $p\gamma$ case with the same input $(\epsilon^{}_+, \epsilon^{}_-) = (1, 1)$. In both cases, one can observe sizable deviations from the prediction of standard neutrino oscillations. Such deviations originate from the sizable values of $\epsilon^{}_\pm$, implying large dissipative effects from Lindblad decoherence. The allowed region shrinks remarkably, which can be clearly seen by comparing the area of golden dots with that of blue dots in the standard case. This is essentially due to the fact that $\Delta^{}_1 - \Delta^{}_2$ is mostly negative, driving the flavor composition $\{f^{\rm D}_e, f^{\rm D}_\mu, f^{\rm D}_\tau\}$ to the democratic value $\{1/3, 1/3, 1/3\}$. Then, another set of parameters $(\epsilon^{}_+, \epsilon^{}_-) = (1, 0)$ is considered in the $p\gamma$ case in order to show the impact of no cancellation between the contributions from neutrinos and antineutrinos. The allowed region of $\{f^{\rm D}_e, f^{\rm D}_\mu, f^{\rm D}_\tau\}$ in this case has been plotted as green dots in the equilateral triangle in Fig.~\ref{fig:hexagon}, where one can observe that the area has been shifted towards smaller values of $f^{\rm D}_e$.

The neutrino fractions $\{x^{\rm D}_e(pp), x^{\rm D}_\mu(pp), x^{\rm D}_\tau(pp)\}$ in the $pp$ case with $(\epsilon^{}_+, \epsilon^{}_-) = (1, 1)$ are plotted as unfilled circles in the isosceles triangles of the flavor hexagon and labelled by ``L-pp" (only explicitly for $x^{\rm D}_e$). Because of $\epsilon^{}_+ = \epsilon^{}_- = 1$, the values of $\{x^{\rm D}_e(pp), x^{\rm D}_\mu(pp), x^{\rm D}_\tau(pp)\}$ turn out to be exactly the same as those in the standard case. This can be clearly seen from Eqs.~(\ref{eq:xeppLind})-(\ref{eq:xtppLind}), which will be reduced to $x^{\rm D}_e(pp) = x^{\rm D}_\mu(pp) = x^{\rm D}_\tau(pp) = 1/2$ for $\epsilon^{}_+ = \epsilon^{}_-$. In the $p\gamma$ case $(\epsilon^{}_+, \epsilon^{}_-) = (1, 0)$, the allowed values of $\{x^{\rm D}_e(p\gamma), x^{\rm D}_\mu(p\gamma), x^{\rm D}_\tau(p\gamma)\}$ have been plotted as green, blue and pink ``$\times$" points, respectively, and labelled by ``L-p$\gamma$" (only explicitly for $x^{\rm D}_e$). All these values $x^{\rm D}_e(p\gamma) \in [0.728, 0.825]$, $x^{\rm D}_\mu(p\gamma) \in [0.575, 0.648]$ and $x^{\rm D}_\tau(p\gamma) \in [0.607, 0.654]$ are not significantly different from those in the standard picture of neutrino oscillations. The approximate $\mu$-$\tau$ symmetry in both scenarios of the standard oscillations and the Lindblad decoherence manifests itself as the overlap of the $\tau$-isosceles triangle after the rotation by an angle of $4\pi/3$ around the center of the hexagon with the $\mu$-isosceles triangle.

It should be noticed that the initial flavor composition of UHE neutrinos and antineutrinos is yet to be measured in neutrino telescopes, the possibilities other than those given in Eqs.~(\ref{eq:ppratio}) and (\ref{eq:pgratio}) could lead to very different allowed regions in the flavor hexagon in Fig.~\ref{fig:hexagon}. In addition, for neutrino oscillations in both the standard picture and the presence of Lindblad decoherence, the total flux of neutrinos or antineutrinos of three flavors is conserved. However, in the scenario of neutrino decays, the situation will be quite different~\cite{Beacom:2002vi}. For a brief comparison, we assume that only the light neutrino mass eigenstate $\nu^{}_1$ is stable, and the heavier ones will decay away completely with invisible decay products. In this scenario, the flavor ratio of UHE neutrinos and antineutrinos is given by $f^{\rm D}_e : f^{\rm D}_\mu : f^{\rm D}_\tau = |U^{}_{e1}|^2 : |U^{}_{\mu 1}|^2 : |U^{}_{\tau 1}|^2$, which has been shown as red dots within the equilateral triangle in Fig.~\ref{fig:hexagon}. One can observe that the area of these red dots is well separated from and significantly larger than that in the case of standard neutrino oscillations. Moreover, the neutrino fractions $\{x^{\rm D}_e, x^{\rm D}_\mu, x^{\rm D}_\tau\}$ are equal to $1/2$ for the $pp$ collisions at the source, and to $(1 - |U^{}_{\tau 1}|^2)/(1 + \Delta^{}_1)$ for the $p\gamma$ collisions. In the latter case, $x^{\rm D}_\alpha \in [0.770, 0.933]$ for $\alpha =e, \mu, \tau$ have been shown as cyan, yellow and black circles in Fig.~\ref{fig:hexagon}. Similarly, one can also present the flavor composition of UHE neutrinos and antineutrinos in other new physics scenarios by using the flavor hexagon.

\section{Summary}\label{sec:summary}

Motivated by recent progress in the detection of UHE neutrinos at IceCube and particularly by the candidate event for the Glashow resonance, we propose a hexagonal parametrization of the flavor composition for both neutrinos and antineutrinos. Such a geometrical description will be useful for the presentation of future experimental results on the detection of both UHE neutrinos and antineutrinos, and also for phenomenological studies of standard and nonstandard particle physics that may result in very different flavor compositions between neutrinos and antineutrinos.

As illustrative examples, two representative scenarios of UHE neutrino flavor conversions have been considered. First, the standard picture of three-flavor neutrino oscillations is reexamined with the latest results of neutrino mixing parameters from the global-fit analysis of all neutrino oscillation data. Assuming the initial flavor composition from either $pp$ or $p\gamma$ collisions as the origin of UHE neutrinos, we derive the exact analytical formulas of the flavor compositions $\{f^{\rm D}_e, f^{\rm D}_\mu, f^{\rm D}_\tau\}$ and the neutrino fractions $\{x^{\rm D}_e, x^{\rm D}_\mu, x^{\rm D}_\tau\}$ at the detector. Then, the scenario of neutrino oscillations in the presence of Lindblad decoherence is discussed. The main motivation for such a scenario is that UHE neutrinos and antineutrinos may experience different interactions with the environments in the production regions. The analytical expressions of the flavor composition in this scenario have been obtained as well. In both scenarios, the hexagonal description has been applied to present the numerical results.

In the near future, we really expect an exciting discovery of the Glashow resonance at IceCube and precision measurements at its successors. Then the determination of neutrino and antineutrino flavor composition begins to be important in diagnosing the production mechanism of UHE neutrinos and exploring the mystery of cosmic accelerators. We hope the proposed hexagonal plot will prove to be practically useful in this exploration.

\section*{Acknowledgements}

The author thanks Dr. Guo-yuan Huang and Prof. Tommy Ohlsson for helpful discussions, and Prof. Zhi-zhong Xing for valuable comments and suggestions. This work was supported in part by the National Natural Science Foundation of China under Grant No.~11775232 and No.~11835013, and by the CAS Center for Excellence in Particle Physics.

\end{document}